**Photonic stopband and reflectance asymmetry from thickness gradient in opals**

Alex Grant[1] and Colm O'Dwyer[1,2,3,4]*

[1] *School of Chemistry, University College Cork, Cork, T12 YN60, Ireland*
[2] *Tyndall National Institute, Lee Maltings, Cork, T12 R5CP, Ireland*
[3] *AMBER@CRANN, Trinity College Dublin, Dublin 2, Ireland*
[4] *Environmental Research Institute, University College Cork, Lee Road, Cork T23 XE10, Ireland*

**Abstract**

The influence of thickness gradient and structural order on the spectral response of opal photonic crystals (PhCs) grown by evaporation-induced self-assembly (EISA) are presented. SEM imaging and angle resolved optical transmission spectroscopy are used to investigate the evolution of the PBG along a thickness gradient for opals grown from five different colloidal sphere concentrations at two different evaporation rates. The degradation of structural order along the thickness gradient is demonstrated, the occurrence of which attenuates the PBG with the thinning of the opal film and results in asymmetrical angle-resolved transmission spectra. The asymmetry in transmission intensity becomes more pronounced for opals grown from lower volume fractions, where secondary Bragg reflections also appear at low incident angles.

**Introduction**

Opals are a type of photonic crystal (PhC) which exhibit striking colors determined by their structure. The optical response of PhCs originates from the periodic refractive index contrast within the crystal, which can be implemented by using materials of alternating refractive index.(1-3) Further tuning of the optical response of these PhCs can be realised by forming structures with graded refractive index.(4) The contrast depends on both the materials used, and the path length which light propagates through each medium. The graded index can be designed by changing the thickness of the individual layers or the dimension of the unit cell from which the crystal is constructed.(5-7) Our recent work investigates how opal thickness can be controlled using volume fraction and evaporation rate and the impact of thickness on their associated optical fingerprints.(8) The observed colours are a result of the wavelengths reflected from each crystal, caused by an interplay between coherent Bragg diffraction and incoherent diffuse scattering that form the photonic band gap (PBG).(9, 10) Extensive reviews of the characteristics which underpin these structures, and their applications are provided elsewhere.(11, 12)

In top-down fabrication methods where compounds are formed layer by layer, these graded thickness PhCs can be designed easily in 1D, 2D and 3D.(13-15) In this case, the thickness gradient is parallel to the growth direction that induces a graded refractive index contrast which shifts the position of the PBG. The value of such materials has been demonstrated in applications such as waveguides,(16, 17) biosensors,(18) optical switches and filters,(19, 20) along with templates for microporous and macroporous materials.(21) In these applications, the benefit of the thickness gradient is primarily in its consequence on the optical response, where the frequency of the photonic stopbands associated with the crystal is tuned based on thickness.

Bottom-up self-assembly methods are simple and cost-effective in the formation of PhCs. However, structural control is difficult, given that assembly proceeds spontaneously. Here, we show that opal PhCs formed by EISA demonstrate a natural thickness gradient, analogous to the "coffee-ring" effect.(22) The gradient in this case is perpendicular to the growth direction, following the direction of evaporation flux.

The result is regions of uniform localized thickness parallel to the growth direction. As a result, spectra acquired at any point are formed from light propagating through uniform thickness, so the gradient does not affect the frequency of the PBG. Instead, we investigate the effects of the thickness gradient on intensity of the transmitted light. We show that the thickness gradient results in a predictable optical response, and angle-resolved transmission spectra are used in tandem with cross-sectional SEM imaging to map the effects of the gradient on the optical fingerprint of opals grown under a range of volume fractions at two different evaporation rates. Furthermore, SEM imaging shows that structural order is progressively reduced along the gradient, which attenuates the photonic band gap (PBG). 3D transmission spectra provide unique optical fingerprints for opals PhCs grown under a range of conditions, all of which demonstrate an asymmetry caused by attenuation of the PBG at regions of least thickness.

## Experimental

### Materials and Substrate Preparation

Colloidal solutions of polystyrene (PS) spheres in 2.5 % (w/v) aqueous suspension with a nominal diameter of 370 nm were purchased from Polysciences Inc. The sphere suspension possessed a net negative charge due to functionalization using a sulfate ester during formation to aid in self-assembly. The substrates used were fluorine-doped tin oxide (FTO)-coated soda-lime glass purchased from Solaronix SA. Glass of thickness 2.2 mm was cut into pieces with dimensions 10 × 25 mm$^2$. The conductive layer of the pieces consisted of a thin FTO layer followed by a thin SiO$_2$ layer (each layer ~20 nm) followed by a thicker FTO layer (~300 nm), all deposited on a borosilicate glass substrate. All substrates were cleaned using the same process: Successive sonication in acetone (reagent grade 99.5%; Sigma Aldrich), isopropyl alcohol (reagent grade 99.5%; Sigma Aldrich) and de-ionized water, each for 5 minutes. All cleaned samples were dried at room temperature in an inert atmosphere. Kapton tape was used to cover the back of the substrate and partially cover the surface of the substrate such that an exposed front surface of 10 × 10 mm$^2$ remained. This ensured colloidal crystal formation occurred only on the desired surface area. The samples were treated in a Novascan PSD Pro Series digital UV-ozone system for 1 h to maximize hydrophilicity of the surface of the substrates immediately prior to photonic crystal formation.

### Evaporation-Induced Self-Assembly

Opals were formed using evaporation-induced self-assembly. FTO glass substrates were immersed in a diluted PS sphere solution with volume fractions in the range of 0.025% to 0.125 % which were placed on a hot plate at a constant temperature between 28 °C and 34 °C. Once the sphere solution was evaporated fully, the substrates were removed and left to dry in inert atmosphere at room temperature. To ensure consistent EISA of the colloidal films at a well-defined angle, we used Vat polymerization 3D printing to create beakers that were formed from photopolymerized poly(methyl methacrylate)-based resin using a Formlabs Form 2 stereolithography printer. All beakers were cleaned using the same sonication process described above, followed by UV-ozone treatment for 1 h. The diameter of each beaker was 6 cm and substrate rests were positioned at an angle of 30° to the beaker walls.

### Microscopy and Spectroscopy

All microscopy was performed using a FEI Quanta 650 SEM at accelerating voltages of 10 kV – 30 kV. Prior to imaging, an incision was made along the length of each sample using a thin blade so that the cross-section could be imaged for thickness measurements. Sputter coating on non-conductive PS spheres was carried out using a Quorum 150T S magnetron sputtering system. Sputtering samples reduced charging effects, namely electron diffusion and beam drift, allowing better contrast and feature definition. For a thickness line measurement $L$, the actual thickness is calculated according to $T = L/\cos 45° = L\sqrt{2}$. Thickness measurements were enabled by tilting the sample stage parallel to the surface at an angle of 45°. For each sample, several thickness measurements were obtained and the mean values in each case used as data points to establish the volume fraction-thickness relationship.

Angle-resolved transmission spectra were used to characterise colloidal opal films. The light source was a tungsten-halogen lamp with an operating wavelength range of 400 to 2200 nm. Transmission measurements were recorded using a UV-visible spectrometer (USB2000+ VIS-NIR-ES) with an operational range of 350 to 1000 nm and a NIR spectrometer (NIRQuest512-2.5) with an operational range of 900 nm to 2500 nm, both from Ocean Optics Inc. Spectra were acquired for an incident angle range of -25° to 25° with a full spectrum recorded for each 1° increment in angle. The integration time was 100 s averaged over 25 scans. Adjustments to the rotation of the sample were made using a motorized rotational stage (ELL8; Thorlabs Inc.) which rotated the sample while the incident beam and detector arm remained fixed. For all measurements, the sample was placed on its side (25 mm$^2$ side) and angle displacement occurred perpendicular to the direction of the thickness gradient, opposite to the direction of evaporation flux during growth, corresponding to a small region at either side of the normal.

**Results and Discussion**

The central wavelength of a photonic stopband can be calculated using the Bragg-Snell law:(23)

$$\lambda_{hkl} = \frac{2d_{hkl}}{m}\sqrt{n_{eff}^2 - \sin^2\theta} \qquad (1)$$

The PBG is the primary stopband of an opal and is formed due to Bragg reflections from the (111) plane of a face-centred cubic (fcc) lattice. The interplanar spacing, $d_{111}$ is between adjacent (111) planes and can be calculated as 0.816$D$, where $D$ is the average opal sphere diameter. The effective refractive index $n_{eff}$ can be calculated using the Drude model for a two-material system and $\theta$ is the incident angle relative to the (111) fcc lattice plane.(24) First order resonance is assumed ($m$ = 1). A perfectly ordered opal consists of layers of spheres stacked in the [111] direction of an fcc lattice. For a given incident angle, the central wavelength of the photonic stopband is dependent only on interplanar spacing and refractive index contrast.

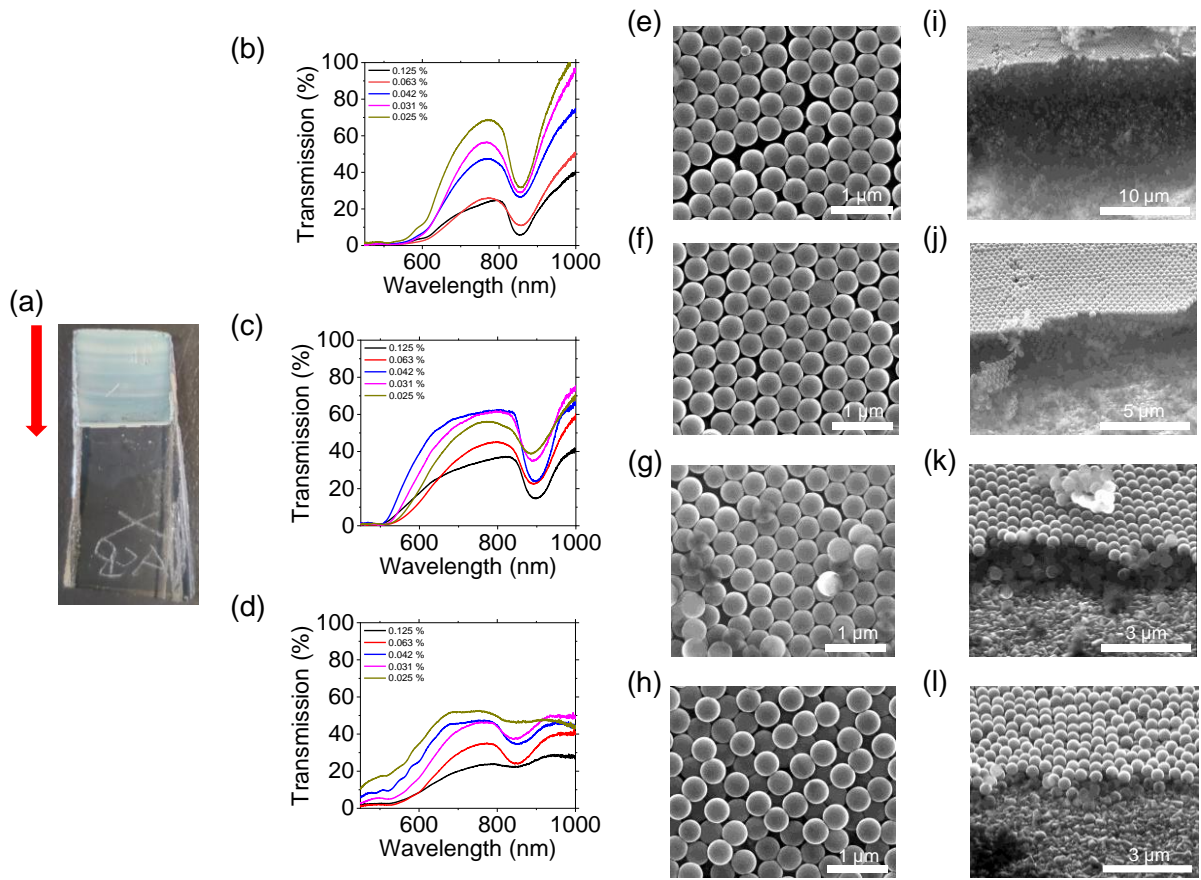

**Figure 1.** Optical and structural characterization of opals grown by EISA at an evaporation temperature of 34 ºC. (a) Photograph of opal sample. Opal thickness decreases in the direction of the red arrow. Optical transmission spectra for five opals, denoted by their volume fractions from 0.025 % to 0.125 %. (b) Spectra obtained at an angle of -25⁰, corresponding to a location near the tail of the red arrow. (c) Spectra obtained at normal incidence, near the centre of the sample. (d) Spectra obtained at an angle of 25⁰, at a location near the head of the red arrow. (e)-(g) SEM images obtained normal to the stacking direction at four positions along the direction of the red arrow shown in (a). (i)-(l) Corresponding cross-sectional SEM images obtained at four positions along the direction of the red arrow in (a) captured at an angle of 45⁰ to the stacking direction.

For all opals examined here, the average sphere diameter is ~400 nm and remains relatively constant. Average sphere diameters for each opal were calculated from over 100 measurements and the maximum standard deviation for any opal examined is 14 nm. Therefore, any shift in the PBG is caused primarily by the effective refractive index, which is dependent on the constituent media within the material and the stacking arrangement of these media. All opals here consist of polystyrene spheres and the air which fills the voids between them. Therefore, PBG shift can occur if the stacking arrangement deviates from a perfectly ordered fcc lattice stacked normal to the (111) plane. Its position is unaffected by crystal thickness.

Figure 1 illustrates the effect of structural order and thickness on the optical transmission spectra for opals grown from a range of volume fractions by EISA at an evaporation temperature of 34 ºC. The photograph in Figure 1 (a) shows the opal from which the SEM images were acquired. The red arrow indicates the direction along which angle-resolved transmission spectra were acquired. The incident angle range chosen was from -25º, corresponding to the edge of the sample near the tail of the red arrow, to 25º, corresponding to a location near the head of the red arrow. Incident angles are relative to the normal to the (111) fcc lattice plane.

Figure 1 (b), (c), and (d) were acquired for opals of five volume fractions at an angle of -25º, 0º, and 25º, respectively. Our recent work investigates the evolution of the optical spectra for opals grown from these five volume fractions at normal incidence. Crystal thickness increases with volume fraction. For the investigated volume fraction range investigated of 0.025 % to 0.125 %, crystal thickness from 4.6 µm (9 layers) to 16.3 µm (33 layers). Moving from high to low volume fractions, we observe in Figure 1 some key changes to the PBG. As volume fraction and hence thickness decreases, the PBG is attenuated and its depth is progressively reduced, transmission at the centre decreases and the band broadens. Due to the thickness gradient, the same changes occur for each given volume fraction as spectra are acquired along this direction, evident in the spectra for at each of the three locations shown here.

There is little observable change in the central wavelength of the PBG between all volume fractions in Figure 1 (b) and (c), indicating a high degree of order at these locations. In Figure 1 (d), the central wavelength of the PBG varies, indicative of a higher concentration of disordered sites. In Figures 1(e)-(h), plan view SEM images are presented in the order of the locations at which they were acquired along the red arrow. The cross-sectional SEM images in Figure 1 (i)-(l) are presented in the same order. Both order and thickness of the opal decrease in this direction along the sample. The plan view SEM images show that there is a significant difference in stacking order along the direction of the red arrow, opposite to the direction of evaporation flux. While in Figures 1 (e) and (f), there is clear fcc symmetry with some disordered domains, Figures 1 (g)–(h) show more disorder with numerous defects and non-close-packed arrays. The thinning of the opal film toward the evaporation line is illustrated by the cross-sectional SEM images in Figures 1 (i)–(l) from a maximum of over 30 layers to a single layer.

Transmitted intensity exponentially decays with crystal thickness.(25) The transmission spectra measured here for all volume fractions show an increase in transmission intensity across the investigated spectral region at locations along the thickness gradient toward the terminating evaporation line. In transmission spectra, disorder presents as increased transmission around the stopband frequencies, with a background of decreased transmission for frequencies outside the stopband, resulting in an attenuation of the PBG.(26) The PBG is well defined and of greatest depth at the region of greatest thickness, while the PBG is attenuated for all volume fractions for the three sets of spectra obtained along the negative thickness gradient. The increase in disorder this direction is shown clearly by the plan view SEM images.

Figure 2 shows angle-resolved transmission spectra for a range of incident angles from -25 º to 25º, relative to the normal to the plane of incidence for opals grown at an evaporation temperature of 34 ºC. Even though just a few degrees higher in temperature, this has been shown to have a dramatic effect on the rate of evaporation and opal formation by EISA. The spectra from these samples are presented in order of decreasing volume fraction (and hence thickness) from Figures 2 (a)–(e). Considering the dispersion relation for the (111) Bragg reflection for an fcc lattice, the opals are well-behaved for all volume fractions at this evaporation rate. The PBG is easily identified by the overhead colormaps as a thin band of lower transmission intensity towards the red edge of the investigated spectral region with a central wavelength at normal incidence occurring from 893–901 nm. The most notable aspects of these spectra are in the asymmetry in both transmission intensity and width of the PBG which is caused by the thickness gradient of the samples.

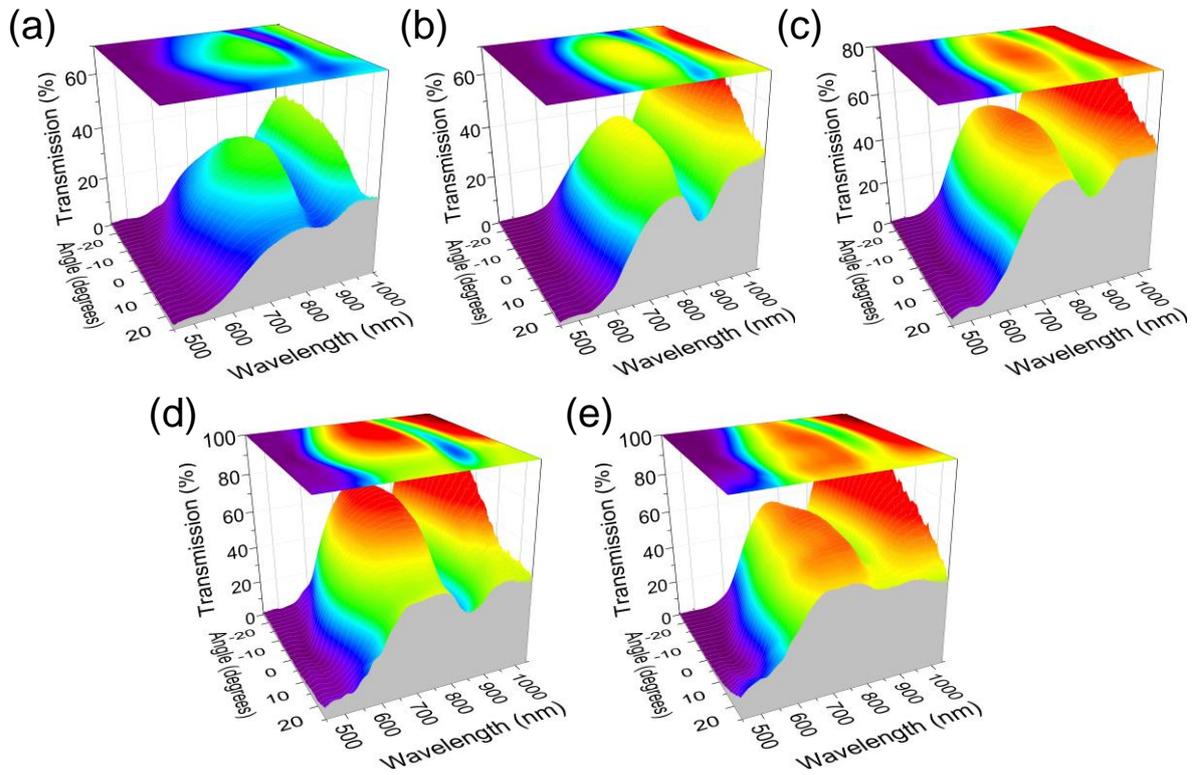

**Figure 2.** Angle-resolved transmission spectra showing the variation of optical transmission as a function of angle of incidence and wavelength, for opals grown by EISA at 34 °C. The spectra correspond to transmission intensity for a range of volume fractions of (a) 0.125 %, (b) 0.063 %, (c) 0.042 %, (d) 0.031 %, and (e) 0.025 % at a range of incident angles from -25° to 25°. Corresponding 2D color maps overhead show the degree of symmetry in the angle-resolved response of EISA grown opals. The color palette is linked to transmission intensity.

In angle-resolved transmission spectra, the curvature of the bands associated with the PBG are important and this contrast is linked to the degree of order. Based on the Bragg-Snell law, the frequency of the PBG changes with incident angle and higher angles result in a blue shift of the PBG. In the angle-resolved spectra for these five volume fractions the PBG central wavelength is blue-shifted at -25°, where thickness and order are maximum, to between 853 nm and 858 nm, which is in good agreement with the Bragg-Snell model. At +25° the PBG central wavelength is further shifted to between 845 nm and 849 nm, which is not in agreement with the Bragg-Snell model, and this deviation is caused by the gradient in thickness and order. The optical response due to the thickness gradient can be clearly observed by the change in color maps for each volume fraction. The band associated with the PBG increases from the low intensity blue color to green from Figure 2 (a)-(e), while the frequencies in the vicinity of the PBG increase from green to red color. Increased transmission is a direct result of reduced close-packed order and thinner opal film regions.

The optical response in opals tends to change at high incident angles, and the differences tend to be most pronounced in the high-energy region ($\sqrt{2}D/\lambda > 1$) where Bragg diffraction by several crystallographic plane families occurs.(27-29) Under this range of conditions, secondary transmission dips appear for high negative incident angles (-25° to 10°) at locations where thickness is minimal. In our case, the secondary transmission dip observed at ~517 nm at -25° and is associated with the (220) Bragg reflection of an FCC lattice.(30) While some of the other Bragg reflections associated with transmission in opals only occur at incident angles outside the range investigated in this work, there is an absence of any transmission dips associated with reflection from the $(11\bar{1})$ and (200) planes, which are expected to appear at energies between those of the (111) and (220) bands.(31-33) This secondary stopband produces the observed blue-green color of the opal samples, as shown in the photograph in Figure 1 (a).

Angle-resolved transmission spectra of opal PhCs performed in other studies have shown that these materials exhibit a highly symmetrical spectral fingerprint over a broad incident angle range up to ±45° due to their thickness uniformity.(34) Elsewhere, the evolution of the PBG and overall spectral response for samples with two uniform thicknesses of 15 μm and 25 μm using angle-resolved reflectance spectra over an incident angle range of ±60⁰ were interrogated.(35) Similar effects were evident, with an attenuation of the PBG and increased reflectance at its edges along with greater reflectance at the higher energy side of the spectra due to interactions with lattice planes other than the [111] family.

The typical approach used to acquire angle-resolved transmission/reflectance measurements is to change the incident angle by rotation of the incident beam or detector arm such that all measurements are obtained at a fixed point on the sample. Here, it is the sample that is rotated using the motorised rotation stage such that the angle-resolved transmission measurements interrogate a small region on either side of the absolute normal. Accordingly, the beam interaction area on the sample moves acutely along the direction in which both thickness and order decrease. The asymmetry in spectral intensity arises from the attenuation of the PBG along this direction. Given that the incident angle increases along the thickness gradient towards the thinner regions, the asymmetry is always observed with lower transmission and well defined spectral features on the negative side of the normal (where the thickness and order are reduced), and an attenuation of the same spectral features on the positive side.

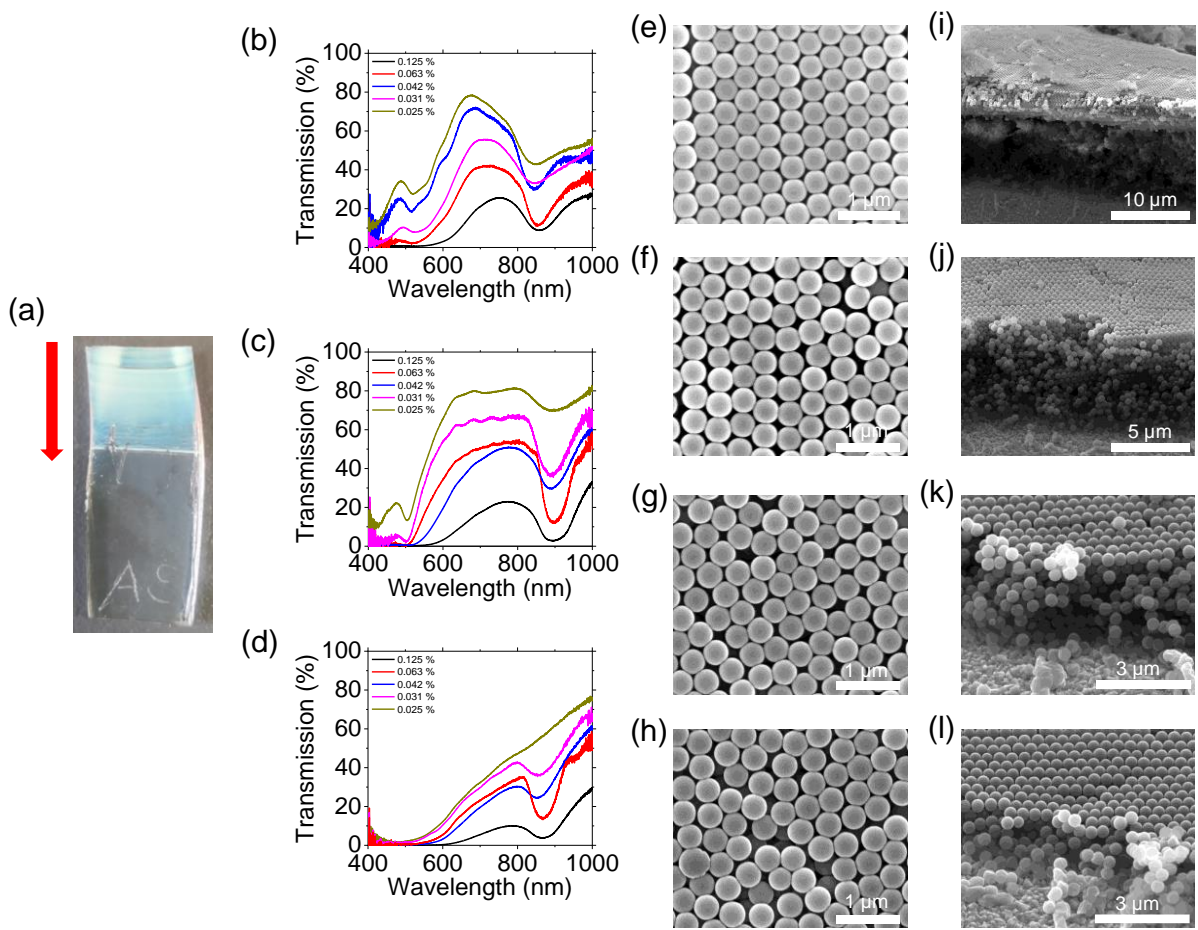

**Figure 3.** Optical and structural characterization of opals grown by EISA at an evaporation temperature of 28 ⁰C. (a) Photograph of opal sample. Opal thickness decreases in the direction of the red arrow. Optical transmission spectra for five opals, denoted by their volume fractions from 0.025 % to 0.125 %. (b) Spectra obtained at a sample angle of -25⁰, corresponding to a location near the tail of the arrow. (c) Spectra obtained at normal incidence, near the centre of the sample. (d) Spectra obtained at an angle of 25⁰, at a location near the head of the arrow. (e)-(g) SEM images obtained normal to the stacking direction at four positions along the direction of the red arrow shown in (a). (i)-(l) Cross-sectional SEM images obtained at four positions along the direction of the red arrow in (a) captured at an angle of 45⁰ to the stacking direction.

Figure 3 shows the optical and structural characterisation of opals grown at an evaporation temperature of 28 ⁰C, corresponding to an evaporation time measured to be twice as long as the previous evaporation temperature of 34 ⁰C. Figure 3 (a) shows a photograph of a representative opal, with the observed blue-green color arising from the secondary stopband. The transmission spectra for opals grown from all five volume fractions obtained at incident angles of -25º, 0º, and 25º are shown in Figure 3 (b), (c), and (d), respectively. For the same range of volume fractions from 0.025 % to 0.125 %, crystal thickness increases from 7.5 µm (15 layers) to 17.5 µm (37 layers). The spectra are notably different from their counterparts grown at 34 ºC. At normal incidence, the low energy side of the spectrum follows the expected shape of a well-defined PBG and a transmission intensity which decreases as a function of volume fraction and hence thickness. At the high negative incidence of -25º, the PBG is blue-shifted as expected based on the Bragg-Snell law. Like the opals grown at 34 ºC, secondary transmission dips associated with the (220) reflection from an fcc lattice occur at ~517 nm at -25º. In this case these (220) dips are present and most well-defined at normal incidence, occurring at ~503 nm. While the PBG is blue shifted by incident angle based on Bragg-Snell law for reflection from a (111) fcc lattice plane, the secondary reflection is red-shifted given that the frequency of the secondary stopband occurs due to reflection from the (220) plane. The appearance of the stopband from -25º to normal incidence indicates that at an evaporation temperature of 28 ºC, there is a high concentration of stacking faults in the form planes orientated in the (220) direction. At 25º, the spectra correspond to the location of least thickness and the (220) dips are absent. The combination of increased disorder and minimum thickness severely attenuates the PBG and suppresses transmission in the blue region of the spectrum.

There is a significant increase in transmission between the primary and secondary transmission dips, where Fabry-Pérot resonances in ordered fcc assemblies are normally found, and we see that the edges of the PBG are attenuated on the lower energy red side resulting in a highly asymmetric PBG. The spectra obtained at high positive incidence of 25º show a well-defined PBG at the two highest volume fractions. As shown by our previous work, the PBG is highly attenuated at the three lowest volume fractions at this evaporation temperature, leading to the conclusion that this evaporation rate-volume fraction combination is the limit below which a breakdown in fcc symmetry for opals grown by EISA occurs. For opals grown at 34 ºC, for the spectra acquired at the thickest region at higher negative incidence, there is high transmittance at the red side of the PBG, resulting in a red-skewed PBG for all volume fractions. In the 28 ºC case, the opposite occurs and a blue-skewed PBG occurs, particularly for the two lowest volume fractions. It appears that these crystals are better able to transmit blue light. Diffuse transmittance is also evident between the PBG and the secondary stopband, presenting in the form of Fabry-Perot resonances.(36)

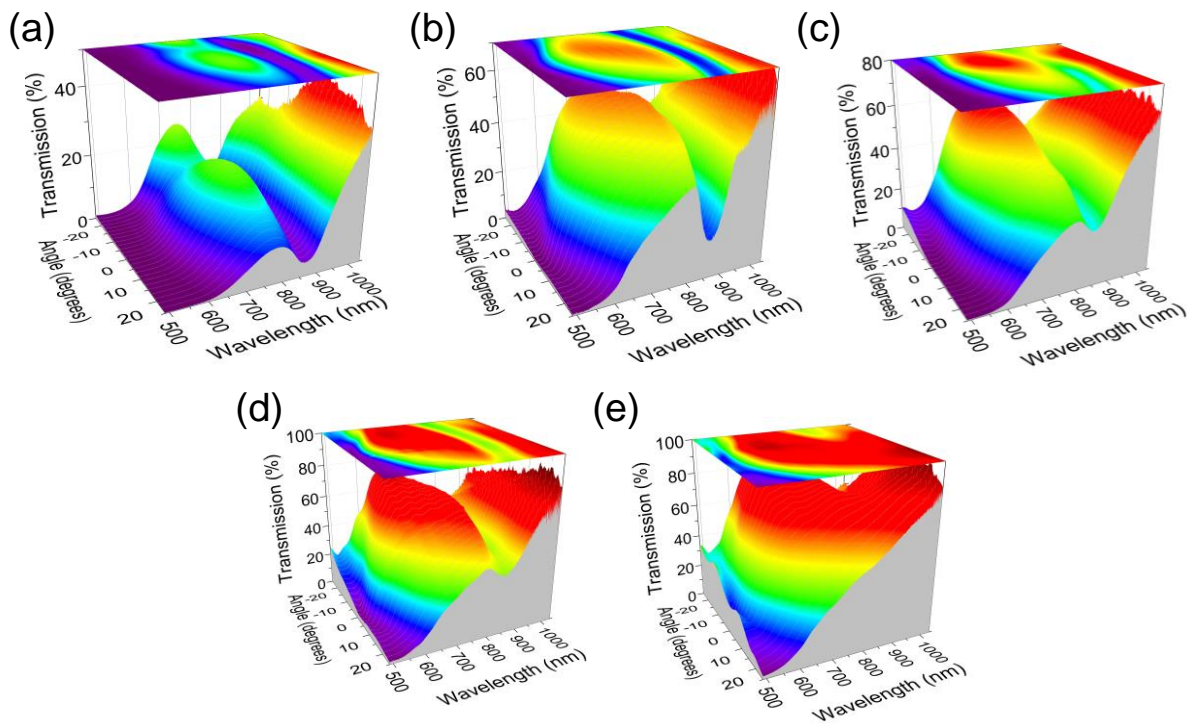

**Figure 4.** Angle-resolved transmission spectra of opals grown by EISA at 28 ºC for a range of concentrations of (a) 0.125 %, (b) 0.063 %, (c) 0.042 %, (d) 0.031 %, and (e) 0.025 % at a range of incident angles from -25º to 25º with corresponding 2D color maps of transmission intensity overhead.

Figures 4 (a)-(e) show the angle-resolved transmission spectra in order of increasing volume fraction from 0.025 % to 0.125 %. The fwhm of the PBG is broadest and of lowest transmission intensity at the largest volume fraction of 0.125 %. Transmission at the PBG increases with volume fraction as expected. Like the 34 ºC case, there is an asymmetry in the spectral window at the PBG, which becomes particularly pronounced at lower volume fractions. Unlike the 34 ºC case, the complete breakdown of fcc symmetry at the lowest volume fraction at higher positive angles of incidence is evident in Figure 4 (e), particularly in the 2D colour map which is saturated fully in red color (higher light transmission) as the PBG band disappears for all angles except those of highest negative incidence angles.

The secondary transmission dip corresponding to the (220) reflection of the fcc lattice is most evident from -25º to 0º for the two lowest volume fractions. The (220) dip can also be distinguished from -25º to 10º at the median volume fraction of 0.042 % in Figure 4 (c). In all cases the gradient to thinner opal films influences transmission across the full spectral region. Nevertheless, the spectral shape of the PBG is well defined for all except the three lowest volume fractions. Even in the unfavourable conditions, larger thickness does not attenuate the PBG, and the PBG is best defined in the regions of maximum thickness. It is the decrease in structural order that causes attenuation and the regions of low thickness tend to form non-close packed arrays, defective areas and dislocations, which significantly disrupt their associated spectral fingerprints.

**Conclusions**

Opals grown here by EISA demonstrate a thickness gradient along which stacking order decreases. This results in a reduction of the contrast and reflectivity of the PBG and is observed within the interaction beam waist of an opal film that is rotated between -25 and + 25º about the normal. Angle-resolved transmission spectra reveal a clear asymmetry in the spectral response of the PhC and the PBG spectral window. This asymmetry is a result of the thickness gradient and becomes particularly pronounced at low volume fractions of PS sphere used to grow the opal films. While the PBG is known to shift due to thickness at fixed angle, we have shown how the change in transmission intensity occurs in films where the PBG remains fixed in ordered regions and a spectral shift happens in more disordered regions. The influence on the overall spectra is very sensitive to thickness and order changes when the sample film is rotated. The spectra for all opals examined here demonstrate a similar trend, with a clearly defined PBG where thickness and order are greatest, and an attenuated PBG where thickness and order are minimal. Secondary Bragg reflections typically occur at high incident angles but are shown here to appear at low incident angles where growth conditions are unfavourable for highly ordered, thicker opal films.

Changes to thickness and structure are common in many methods of opal photonic crystal growth and this work highlights the care needed in assessing the angular dependence of the PBG either side of the normal and the overall transmission response, particularly when ordered periodicity is critical for the study or application. Changes to thickness and structure that occur at the evaporation line during EISA, or the drying line during dip coating or other coating methods, affect order and this degree of order is shown to be sensitive to evaporation rate and the resulting thickness of the opal film either side of the normal to the incident light.

**Acknowledgements**

We acknowledge support from the Irish Research Council under an Advanced Laureate Award (IRCLA/19/118).